\documentclass[twocolumn,aps,prc,showpacs,showkeys,amsfonts,amssymb,amsmath,superscriptaddress]{revtex4}

\usepackage[english]{babel}

\usepackage{graphicx}
\usepackage{ulem}
\usepackage{cancel}
\usepackage{amsmath}
\usepackage{amssymb}
\usepackage{color}
\definecolor{RED}{rgb}{1.0,0.0,0.0}
\usepackage[usenames,dvipsnames]{xcolor}

\newcommand{\ket}[1]{\mathinner{|{#1}\rangle}}
\newcommand{\bra}[1]{\mathinner{\langle{#1}|}}

\newcommand{\EXP}[1]{\mathrm{e}^{#1}}

\DeclareMathOperator{\re}{Re}
\DeclareMathOperator{\im}{Im}

\newcommand{\imat}{{\mathrm{i}}}
\newcommand{\dmat}{\mathrm{d}}
\newcommand{\Omat}{\mathrm{O}}

\newcommand{\DEF}{\overset{\mathrm{def}}{=}}

\newcommand{\scl}{\underset{\hbar\to0}{\sim}}
                 
\newcommand{\coeffA}{\mathcal{A}}
\newcommand{\coeffB}{B}
\newcommand{\coeffH}{\mathcal{H}}
\newcommand{\coeffT}{\mathcal{T}}
\newcommand{\coeffAT}{\mathcal{A}_{\coeffT}}

\newcommand{\goc}{\mathfrak{c}}

\newcommand{\goin}{\mathrm{in}}
\newcommand{\goou}{\mathrm{out}}

\newcommand{\loopC}{\mathcal{C}}
\newcommand{\loopCtilde}{\tilde{\mathcal{C}}}

\begin{document}

\author{J\'er\'emy Le Deunff}
\affiliation{
   Max-Planck-Institut f{\"u}r Physik komplexer Systeme,
   N{\"o}thnitzer Stra{\ss}e 38,
   01187 Dresden, Germany.
   }
\author{Amaury Mouchet}
\affiliation{
   Laboratoire de Math\'ematiques 
   et Physique Th\'eorique, Universit\'e Fran\c{c}ois Rabelais de Tours 
   --- {\textsc{cnrs (umr 7350)}},
   F\'ed\'eration Denis Poisson,
   Parc de Grandmont, 37200
   Tours,  France.
   }
\author{Peter Schlagheck}
\affiliation{
   D{\'e}partement de Physique, University de Liege,
   4000 Li{\`e}ge, Belgium.
   }

\title{Semiclassical description of resonance-assisted tunneling in one-dimensional integrable models}
\date{\today}

\begin{abstract}
Resonance-assisted tunneling is investigated within
the framework of one-dimensional integrable systems.
We present a systematic recipe, based on Hamiltonian normal forms,
to construct one-dimensional integrable models that exhibit resonance 
island chain structures with accurately controlled sizes and positions
of the islands.
Using complex classical trajectories that evolve along
suitably defined paths in the complex time domain, we construct a
semiclassical theory of the resonance-assisted tunneling process.
This semiclassical approach yields a compact analytical expression for 
tunneling-induced level splittings which is found to be in
very good agreement with the exact splittings obtained through 
numerical diagonalisation. 
\end{abstract}


%\keywords{tunneling, dynamical tunneling, resonant tunneling,
%  instantons, semiclassical methods, Wick rotation, complex time,
%  Hamiltonian normal form theory, resonant island chains,
%  resonance-assisted tunneling} 

\maketitle

\section{Introduction}\label{sec:introduction}

In quantum theory, the term \textit{tunneling} defines classically
forbidden processes -- i.e. which cannot be described by real
solutions of Hamilton's equations of motion -- and was
originally employed to characterize transitions which were forbidden by energy
barriers. It has been thereafter extended to \textit{dynamical}
tunneling which refers to any quantum transition between two
classically distinct regions in phase
space~\cite{Davis.Heller_1981a,Maitra.Heller_1996} where the
inhibition of a classical transition between these two regions is
not necessarily restricted to the constraint of energy
conservation.
Indeed, focusing first on the simple case of integrable systems with
one degree of freedom, the development of semiclassical
techniques~\cite{Berry.Mount_1972} have permitted a deeper qualitative
and quantitative understanding of tunneling in terms of classical
trajectories. In particular, its complete semiclassical description
requires, in addition to \textit{real} orbits, to take into account
also \textit{complex} classical
trajectories~\cite{Balian.Bloch_1974,Maitra.Heller_1997}. For
instance, studying scattering phenomena involved in chemical
reactions, Freed~\cite{Freed_1972} and George and
Miller~\cite{George.Miller_1972,Miller_1974} incorporated
complex orbits, evolving along suitable paths in the
complex time domain, in order to compute the Green function
$G(q_f,q_i,E)$ giving rise to the tunneling transmission with an
energy $E$ below the top of a potential barrier.

A few years later, Coleman~\cite{Coleman} (see also
Ref.~\cite{Shifman}) developed an approach suited for the simplest 
bounded systems, where tunneling is generally identified
in the spectrum as small splittings between doublets of nearly
degenerate discrete eigenenergies. In the context of field theories,
he introduced the notion of \textit{instantons} which
corresponds to classical solutions of the Hamilton dynamics once a
Wick rotation $t\rightarrow -\imat t$ has been performed.  
For systems with the standard form of the Hamiltonian
\begin{equation}\label{eq:usual_ham}
H(p,q) = \frac{p^2}{2} + V(q),
\end{equation}
where $p$ and $q$ are the canonical variables, this transformation on
the time leads to an inversion of the potential $V(q) \rightarrow -V(q)$. 
The classical trajectories in the new potential allow to evaluate quantum
observables associated with the lowest energies (ground-state doublet or 
multiplet), such as the frequency of oscillation between an arbitrary 
number of identical minima, or the decay rate of a metastable state 
that is initially defined in a local minimum of the potential
and decays via the coupling to a continuum of unbounded states.

The method has been recently generalized~\cite{LeDeunff.Mouchet_2010},
using again the idea of a suitably parametrized complex time
path, in order to embrace more general situations involving,
e.g., an arbitrary energy and/or Hamiltonians not necessary of the 
form~\eqref{eq:usual_ham}. 
For instance, resonant tunneling, which has been widely investigated 
in one-dimensional (1D) open systems with two consecutive 
barriers~\cite{Bohm,Cruz.Hernandez-Cabrera.ea_1991,Zohta_1990}, is
thus explained in terms of constructively interfering repetitions
of complex orbits. This is shown for the simple case of a
triple-well potential where the presence of a deeper middle well
(which prevents the application of the standard instanton
techniques based on the complete Wick rotation recalled above)
can create giant fluctuations of the tunneling period between the 
two symmetric outer wells~\cite{LeDeunff.Brodier.ea_2012}, namely 
whenever a third energy level, associated with a state that is
localized in this middle well, comes close to a doublet that is
associated with the two outer wells.

If the number of the degrees of freedom exceeds 1, we generically
deal with nonintegrable Hamiltonians whose phase space contains 
regular islands foliated with Kolmogorov-Arnold-Moser (KAM) tori 
and surrounded by chaotic seas. 
In that case, tunneling is drastically modified and yields erratic 
fluctuations (by several orders of magnitude) of the associated rates 
and time scales when varying a parameter of the
system~\cite{Lin.Ballentine_1992,Bohigas.Tomsovic.ea_1993}. 
These fluctuations have the same origin in the quantum spectrum as the
ones observed in the one-dimensional resonant case. 
However, the appearance of natural boundaries~\cite{Greene.Percival_1981}
prevents the analytical continuation of the invariant classical
KAM tori into the complex plane of the classical phase space,
and the methods at work for one-dimensional systems fail. 
Despite some important breakthroughs in that direction during the past few
decades~\cite{Shudo.Ikeda_1995,Shudo.Ishii.ea_2009,Creagh.Whelan_1996},
a full semiclassical description in terms of complex classical
structures is still missing. On the other hand, a considerable effort
has been made to combine, within a  perturbative framework, 
statistical descriptions of chaos-assisted tunneling due to the influence 
of the classical chaotic sea
\cite{Tomsovic.Ullmo_1994,Leyvraz.Ullmo_1996,Creagh.Whelan_1996}
with the theory of resonance-assisted tunneling (RAT) 
\cite{Brodier.Schlagheck.ea_2001,Brodier.Schlagheck.ea_2002,Eltschka.Schlagheck_2005,Mouchet.Eltschka.ea_2006,Lock.Backer.ea_2010,Keshavamurthy.Schlagheck}
that is based on the presence of nonlinear resonances within the
regular regions. This approach has been shown to
provide rather accurate semiclassical predictions of quantum
tunneling rates in kicked model
systems~\cite{Keshavamurthy.Schlagheck}.

In this paper, we present and discuss a semiclassical formula for
resonance-assisted tunneling splittings in one dimensional
integrable systems that exhibit a pair of symmetric regions of 
bounded motion in the classical phase space, each
of them being surrounded by a resonant island chain.
The study of such models is clearly inspired by the recent idea
to mimic regular regions of mixed systems with a fictitious integrable
approximation in order to predict regular-to-chaotic
tunneling~\cite{Backer.Ketzmerick.ea_2008,Backer.Ketzmerick.ea_2010},
although we are not aiming here at approximating a given
nonintegrable system by such a model. Instead, our motivation is to
obtain a fully semiclassical (and nonperturbative) description of
resonance-assisted tunneling through the analytical continuation of
invariant classical manifolds to the complex domain, which is
permitted by the integrability of the Hamiltonian. This will allow
us to understand how the island chains in the phase space are at
work to create fluctuations in the tunneling-induced level
splittings when varying a parameter of the system and what the 
semiclassical conditions are for resonant tunneling.

The paper is organized as follows.  
In Sec.~\ref{sec:normal_form}, we construct a class of
models that fulfill the classical properties mentioned
above using the theory of Hamiltonian normal forms.
Sec.~\ref{sec:splittings} is dedicated to the computation of
tunneling splittings. 
In Sec.~\ref{subsec:semiclassical_formula}, we shall
investigate the complex manifold of our integrable model
and identify the relevant complex classical trajectories, defined along 
well-suited complex time paths, that give rise to a semiclassical formula
[Eq.~\eqref{eq:splitting_cmplx_path} below] for resonance-enhanced
level splittings.
The perturbative RAT method is then applied to our system in
Sec.~\ref{subsec:comp} and compared with the complex paths
approach. We discuss the validity of the two methods in both
limits of small and large sizes of the island chains.

\section{The model}\label{sec:normal_form}

\subsection{Normal form theory}\label{subsec:theory}

The Hamiltonian normal forms in classical 
mechanics~\cite{Arnold,OzoriodeAlmeida} have been originally
developed by Birkhoff~\cite{Birkhoff} and extended by
Gustavson~\cite{Gustavson_1966} with the aim to classify 
the classical dynamics in the neighborhood of the periodic orbits in
nonintegrable systems with several degrees of freedom. This
classification relies on canonical equivalence and provides the
simplest (local) form of the Hamiltonian where the only terms that
are kept are those that are sufficient to supply the intrinsic
``skeleton'' of the dynamics, i.e., those terms that cannot be eliminated
by a canonical transformation because they genuinely encapsulate the
geometrical features of the dynamics. Hamiltonian normal forms have helped
to predict the quantum energy spectra of such 
systems~\cite{Swimm.Delos_1979,Eckhardt_1986}. 

Normal form approaches are based on the combination of Fourier and Taylor 
expansions of the nonintegrable Hamiltonian in the neighbourhood of a 
periodic orbit. 
Order by order, beyond the quadratic terms, a sequence of canonical 
transformations can be explicitly built to eliminate all terms but 
the resonant ones. The latter may give rise to divergencies
manifesting in the above construction procedure, which is
well known as the problem of small denominators. 
Those resonant terms inhibit the integration of systems of ordinary 
differential equations, and generally the procedure to obtain an 
accurate approximation of the dynamics does not converge,
which is a signature of the non integrability of the system. 
Nevertheless, this procedure enables one to extract some essential 
information about the fine structure of the phase space 
as it provides a description not only of the regular part but also of 
the resonant layout where chaos emerges from, thereby leading to the simplest
local integrable approximation of the system.

Specifically, let us consider an autonomous system with two
degrees of freedom (or a periodically time-dependent system 
with one degree of freedom) depending on one control parameter~$\epsilon$.
In the neighborhood of a nondegenerate stable orbit, a transverse
coordinate system~$(p,q)$ in the so-called Poincar\'e surface of
section can be chosen~\cite{Leboeuf.Mouchet_1999} such that the
transverse dynamics is governed by a Hamiltonian whose normal form
is given by (see Ref.~\cite{Leboeuf.Mouchet_1999} for an exhaustive 
and systematic study on this matter)
\begin{multline}\label{eq:gen_normal_form1}
  h^{(\ell)}(p,q;\epsilon) = \frac{\omega(\epsilon)}{2}(p^2+q^2) 
  + \sum_{k=2}^{\lfloor\ell/2\rfloor}{a_k(\epsilon)(p^2+q^2)^k} \\ 
  + b_{\ell}(\epsilon)\re[(p+\imat q)^{\ell}] + \text{higher order terms} \;,
\end{multline}
where $\omega$, $\{a_k\}$ and $b_{\ell}$ are real parameters,
$\lfloor\cdot\rfloor$ denotes the integer part, and the index
$\ell \geq 3$ represents the order of the first angle-dependent resonant 
term occurring in the expansion. 
It can be rewritten in terms of the action-angle variables 
$I = (p^2+q^2)/2$ and $\theta = \tan^{-1}(q/p)$ associated with the 
one-dimensional harmonic oscillator,
\begin{multline}\label{eq:gen_normal_form2}
  \tilde{h}^{(\ell)}(I,\theta;\epsilon) = \omega(\epsilon) I + 
  \sum_{k=2}^{\lfloor\ell/2\rfloor}{a_k(\epsilon) (2I)^k} \\
  + b_{\ell}(\epsilon)(2I)^{\ell/2}\cos{(\ell\theta)} 
  + \text{higher order terms}\;.
\end{multline}
It is straightforward to see that the resonant term
creates, for appropriate values of the parameters,
an island chain of $\ell$ islands in the transverse
dynamics around the origin $(p,q)=(0,0)$ where the periodic orbit of
the 2D system intersects the Poincar\'e surface of
section.

\subsection{The Hamiltonian}\label{subsec:model}

\begin{figure}[t]
\center \includegraphics[width=0.5\textwidth]{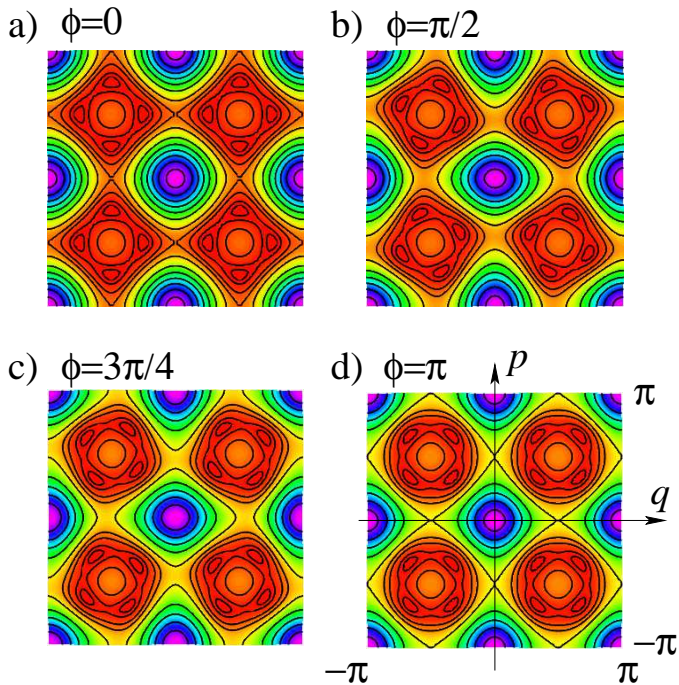}
\caption{\label{fig:ps_period_H4} (Color online) Phase space of the Hamiltonian
\eqref{eq:H4model} with $a_1=1$, $a_2=-0.55$, $|b|=0.05$ for
different values of the phase (a)~$\phi=0$, (b)~$\phi=\pi/4$, 
(c)~$\phi=\pi/2$, (d)~$\phi=\pi$.}
\end{figure}

Let us rewrite  the normal form \eqref{eq:gen_normal_form1} as
\begin{equation}\label{eq:model_gen}
  h^{(\ell)}(p,q) \DEF h^{(\ell)}_0(p^2+q^2) + v^{(\ell)}(p,q)
\end{equation}
with
\begin{subequations}
\begin{align}
  h^{(\ell)}_0(I) \; & \DEF a_1 I + 
  \sum_{k=2}^{\lfloor\ell/2\rfloor} a_k (2I)^k \;, \\
  v^{(\ell)}(p,q) \; & \DEF \re{[b(p+\imat q)^{\ell}]} \;,
\end{align}
\end{subequations}
where the $\{a_k\}$ are real parameters and $b\equiv|b|\exp{(\imat\phi)}$ 
is a complex parameter. From now on, we consider 
$h_{0}^{(\ell)}$ to be the unperturbed part while
$v^{(\ell)}$ has to be understood as a perturbation.
Starting from~$h^{(\ell)}$, we perform the substitution 
$(p,q)\mapsto(\cos{p},\cos{q})$ and thereby obtain the new Hamiltonian,
\begin{equation}\label{eq:model_gen_period}
H^{(\ell)}(p,q) \DEF h^{(\ell)}(\cos{p},\cos{q})\;
\end{equation}
with which we shall work from now on.
This construction gives rise to a smooth periodic replication of the phase
space structure in both position and momentum.
We then restrict $H^{(\ell)}(p,q)$ to the torus 
$[-\pi,\pi]\times[-\pi,\pi]$ by imposing $2\pi$-periodic boundary 
conditions of the wave function in both $p$ and $q$.
For a suitable choice of $\{a_k\}$ and~$b$, this torus encloses
four elementary cells centered at
$(p,q)=(\pm\pi/2,\pm\pi/2)$ and surrounded by a $(\ell:1)$ resonance chain,
as illustrated in Fig.~\ref{fig:ps_period_H4}.
While the modulus $|b|$ controls the size of the island chains, we can 
tune the relative orientation of the main islands by smoothly
rotating the $(\ell:1)$ resonances via the phase $\phi$ (see
Fig.~\ref{fig:ps_period_H4}).

\begin{figure}[t]
\center \includegraphics[width=0.5\textwidth]{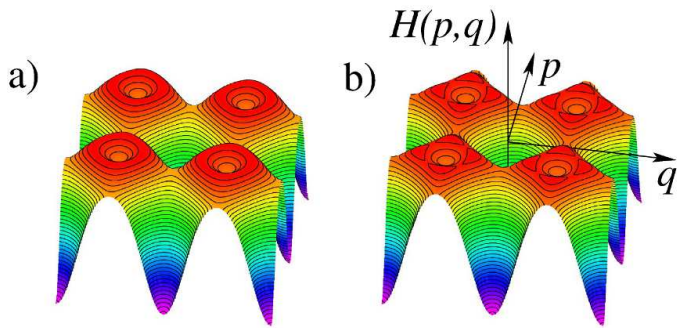}
\caption{\label{fig:volcan} 
(Color online) Graph of $(p,q)\mapsto H(p,q)$ on the fundamental domain 
$[-\pi,\pi]\times[-\pi,\pi]$
where~$H$ is given by Eq.~\eqref{eq:H4model} 
with $a_1=1$, $a_2=-0.55$, $\phi=0$ and (a)~$b=0$ ; 
(b)~$|b|=0.05$.}
\end{figure}

Focusing now on the simplest case $\ell=4$~\footnote{The
normal form of third order exhibits only one chain of three
unstable points as shown in~\cite{Leboeuf.Mouchet_1999}
and therefore is not representative of what occurs for higher
orders.}, we obtain a model with $(4:1)$ resonances:
\begin{eqnarray}\label{eq:model_h4}
 h(p,q) &\DEF& h^{(4)}(p,q)\;, \\
&=& \frac{a_1}{2}(p^2+q^2) + a_2(p^2+q^2)^2 + \re{[b(p+\imat q)^4]}\;. \notag
\end{eqnarray}
The Hamiltonian that we shall work with is given by
\begin{subequations}\label{eq:H4model}
\begin{equation}\label{eq:model_period_H4}
H(p,q) \DEF H^{(4)}(p,q) = H_0(p,q) + V(p,q)\;,
\end{equation}
with
\begin{eqnarray}\label{eq:model_period_detail}
 H_0(p,q) &=& \frac{a_1}{2}(\cos^2p + \cos^2q) + a_2(\cos^2p + \cos^2q)^2\;, \notag \\&&\\
 V(p,q) &=& |b|\left\{(\cos^4p + \cos^4q - 6\cos^2p\cos^2q)\cos\phi \right. \notag \\
 &&\left. \, \, \quad -4(\cos^3p\cos q - \cos^3q\cos p)\sin\phi \right\}\,.
\end{eqnarray}
\end{subequations}
Choosing $a_1>0$ and $a_2<0$, the energy profile $(p,q)\mapsto H(p,q)$
exhibits four symmetric volcano-like patterns within the
torus $[-\pi,\pi]\times[-\pi,\pi]$, each one having a local
minimum located at the center $(\pm\pi/2,\pm\pi/2)$ of the
corresponding elementary cell and four identical maxima
situated along the crown of the volcanos (see Fig.~\ref{fig:volcan}).

\section{tunneling splittings}\label{sec:splittings}

\subsection{Quantum mechanics}

For bounded Hamiltonians $H(p,q)$ with a twofold symmetry, the
spectrum is made of discrete energies $E^{\pm}_{n}$ which can be
classified according to their parity $(\pm)$. They are
determined from the stationary Schr{\"o}dinger equation
\begin{equation}\label{eq:doublet_H}
  H(\hat{p},\hat{q})\ket{\phi^{\pm}_{n}} =
  E^{\pm}_{n}\ket{\phi^{\pm}_{n}}
\end{equation}
(with $\hat{p}$ and $\hat{q}$ denoting the momentum and position operator, 
respectively) where the natural integer $n$ sorts the corresponding 
eigenstates $\ket{\phi^{\pm}_{n}}$ which form an orthonormal basis. 
In the limit $\hbar \rightarrow 0$ (when Planck's constant
is much smaller than the classical phase-space areas), the association of 
the quantum states $\ket{\phi^{\pm}_{n}}$ with classical phase-space 
structures can be visualized using the semiclassical Wigner or Husimi 
distributions~\cite{Husimi_1940,Takahashi.Saito_1985,TorresVega.Frederick_1990}. 
In the case of regular classical dynamics, the states are sharply 
localized [up to order $\Omat(\hbar$)] along invariant tori 
in the classical phase space.

For instance, in the simple case of a 1D Hamiltonian
of the standard form~\eqref{eq:usual_ham} with a potential $V(q)$ that 
exhibits two symmetric local minima, the eigenstates $\ket{\phi^{\pm}_{n}}$ 
with energies $E^{\pm}_{n}$ below the top of the barrier between the
minima are mainly localized on symmetric tori in the wells characterized 
by the classical energy $E_n \simeq E_n^+ \simeq E_n^-$. 
A quantum state that is given by the symmetric or antisymmetric linear 
combination of the eigenstates $\ket{\phi^{\pm}_{n}}$ is no longer 
stationary but gives rise to oscillations from one well to the other with the
period $\tau = 2\pi\hbar/\Delta E_n$, where $\Delta E_n \DEF
|E_n^- - E_n^+|$ is the splitting of the levels $E^{\pm}_{n}$
in the spectrum. This splitting represents, for both integrable
and nonintegrable bounded systems, a characteristic signature of 
tunneling between two classically separated regions in the phase space.

The Hamiltonian~$\hat{H}$ of our quantum model is derived from a
straightforward quantization of the classical
model~\eqref{eq:H4model} with the simple symmetrization rule
\begin{equation}
f(p)g(q) \mapsto \frac{1}{2}\big[ f(\hat{p})g(\hat{q})+g(\hat{q})f(\hat{p})
\big]
\end{equation}
for the product of two functions~$f(p)$ and~$g(q)$ \cite{Weyl,Shewell_1959}.
From the quantization of~\eqref{eq:H4model}, the periodicity of the 
Hamiltonian allows us to use the Floquet-Bloch theorem and to restrict, 
for integer values of $2\pi/\hbar$, the analysis to the finite-dimensional 
Hilbert space $\coeffH$ spanned by strictly periodic eigenstates in both 
position and momentum on the torus $[-\pi,\pi]\times[-\pi,\pi]$.

As we see in Fig.~\ref{fig:ps_period_H4}, two independent twofold 
symmetries are relevant in our model.
It is natural to associate these two symmetries with the antiunitary
operators $\hat{\Pi}_q$ and~$\hat{\Pi}_p$ that perform mirror operations
with respect to the $p$ and $q$ axes, respectively, and that are 
defined through
$\hat{\Pi}_q f(\hat{p},\hat{q}) \hat{\Pi}_q=f(\hat{p},-\hat{q})$
and $\hat{\Pi}_p f(\hat{p},\hat{q}) \hat{\Pi}_p=f(-\hat{p},\hat{q})$
for any function~$f$ of the canonical operators $\hat{p}$ and $\hat{q}$.
Obviously, $\hat{\Pi}_p$~is the standard time-reversal operator, 
while $\hat{\Pi}_q$ is the time-reversal operator composed with the usual 
unitary parity operator.
By construction, $\hat{\Pi}_q$, $\hat{\Pi}_p$, and $\hat{H}$
mutually commute with each other.
However, the time-reversal invariance of the Hamiltonian cannot 
be exploited to discriminate among its eigenstates;
it only allows one to choose the latter to be entirely real.
This particular phase convention fixes the spectrum of~$\hat{\Pi}_q$ 
to be identical to the spectrum of the parity operator, such that we
can classify the eigenstates of the Hamiltonian according to their parity:
$\hat{\Pi}_q\ket{\phi^{\pm}_n} = \pm\ket{\phi^{\pm}_n}$.

In contrast to conventional double-well systems, however, the 
eigenenergies associated with the four main islands within the unit cell
are organized in quartets, and the parity alone is not sufficient to 
unambiguously specify the doublet whose level splitting is determined by
tunneling along, say, the $q$ direction.
To lift this ambiguity, we numerically construct four states 
$\ket{\pm,\pm}$ from the local $n$-th excited harmonic oscillator 
eigenstates $\ket{R or L,U or D}$ that are centered at $(q,p)=(\pm\pi/2,\pm\pi/2)$ 
[with $L$ ($R$) referring to the left (right)
column and $U$ ($D$) to the upper (lower) row within the unit cell
depicted in Fig.~\ref{fig:ps_period_H4}], namely through
\begin{subequations}
\begin{eqnarray}
  \ket{++}&\DEF&\frac{1}{2}\big(\ket{RU}+\ket{LU}+\ket{LD}+\ket{RD}\big)\;;\\
  \ket{-+}&\DEF&\frac{1}{2}\big(\ket{RU}-\ket{LU}-\ket{LD}+\ket{RD}\big)\;;\\
  \ket{+-}&\DEF&\frac{\imat}{2}\big(\ket{RU}+\ket{LU}-\ket{LD}-\ket{RD}\big)\;;\\
  \ket{--}&\DEF&\frac{\imat}{2}\big(\ket{RU}-\ket{LU}+\ket{LD}-\ket{RD}\big)\;.
\end{eqnarray}
\end{subequations}
Being eigenstates of the parity operator (with the eigenvalues $1$
for $\ket{++}$ and $\ket{--}$ and $-1$ for $\ket{+-}$ and $\ket{-+}$)
\footnote{Technically this requires that the dimension~$2\pi/\hbar$ of the 
Hilbert space is an integer multiple of~4. 
Therefore, in the following we will restrict ourselves to the case 
where~$N=\pi/2\hbar$ is integer.},
these states closely mimic the quartet of eigenstates of the Hamiltonian 
that are localized within the centers of the four islands.
In order to focus on tunneling along the $q$ direction, we therefore
select those two eigenstates of $\hat{H}$ that have a 
maximal numerical overlap with~$\ket{++}$ and~$\ket{-+}$. 

Because of the invariance of~$H(\hat{p},\hat{q})$ under the canonical 
transformation~$(\hat{p},\hat{q})\mapsto(\hat{q},-\hat{p})$,
tunneling along the momentum direction will give rise to the same 
splittings as for the position direction.
This equivalence is reflected by an exact degeneracy between two 
energies among the four of the quartet: 
For even $N/4$ the two states having a maximal overlap with~$\ket{+-}$ 
and~$\ket{-+}$ have exactly the same energies, while for odd $N/4$ the 
levels corresponding to~$\ket{++}$ and~$\ket{--}$ are exactly degenerate.

\subsection{Semiclassical theory}\label{subsec:semiclassical_formula}

By construction of the Hamiltonian \eqref{eq:model_period_H4},
the unperturbed case $b=0$ gives rise to a tunneling problem
that is equivalent to the one of a symmetric 1D double-well system.
This scenario has been intensively investigated using JWKB analysis
in order to connect two approximated eigenstates, the so-called quasimodes,
each of them being localized on a distinct real torus
\protect\cite{Wilkinson_1986,Creagh_1997}.
Up to a prefactor of order 1, the level splitting~$\Delta E_n$ 
associated with the doublet at energy~$E_n$ is essentially determined 
as \protect\cite{Landau.Lifshitz,Garg_2000}
\begin{equation}\label{eq:splitting_unpert}
\Delta E_n \scl \hbar \omega_n \EXP{-\Sigma(E_n)/(2\hbar)}\;,
\end{equation}
where $\omega_n$ is the frequency of classical oscillation on the torus 
with energy $E_n$ within the left or right well, and $\Sigma(E_n)$ is the 
imaginary action of a closed complex path that connects the two
symmetric tori. 

The simplest bounded model inducing quantum resonances can be 
obtained~\cite{LeDeunff.Brodier.ea_2012} for a Hamiltonian of the form
\eqref{eq:usual_ham} with a potential~$V$ that has three wells, 
two symmetric outer ones, say, that are separated by a deeper central well. 
In such a system, resonant tunneling arises due to 
the constructive interference of classical paths that are
bouncing back and forth between the two tunneling barriers.
It was shown in Ref.~\cite{LeDeunff.Mouchet_2010} that these relevant orbits 
can be obtained from a suitable complex time path
$s\mapsto \re t(s) +\imat \im t(s)$ where, unlike for a pure Wick 
rotation, both the real and the imaginary part of the complex time are 
necessary to concatenate the primitive orbits that constitute the 
complex trajectories.
For the resonance-assisted tunneling problem presented by the
model~\eqref{eq:H4model} with $b\neq 0$, we shall, in the same spirit, 
introduce a generic type of concatenated complex paths that connect two 
real symmetric tori inside the main island of each cell. 
This set of orbits will be used to predict tunneling splittings 
between states that are localized within these islands. 
For a given real parametrization $s \mapsto t(s)$ of a complex time path, 
it can be easily shown \cite[\S~III.A]{LeDeunff.Mouchet_2010}, using 
the analyticity of the Hamiltonian~$H$ with respect to $(p,q,t)$, 
that the complex Hamiltonian equations 
\begin{subequations}
\begin{align}
&\frac{\dmat p}{\dmat s} = -\frac{\partial H}{\partial q} \frac{\dmat t}{\dmat s}\;; \label{eq:cmplx_ham_p}\\
&\frac{\dmat q}{\dmat s} = \phantom{-}\frac{\partial H}{\partial p} \frac{\dmat t}{\dmat s} \label{eq:cmplx_ham_q}
\end{align}
\end{subequations}
are equivalent to the set of real Hamiltonian equations
\begin{subequations}\label{eq:real_ham}
\begin{align}
&\frac{\dmat (\re q)}{\dmat s} = \phantom{-}\frac{\partial}{\partial (\re p)} \left[\re \left(H\frac{\dmat t}{\dmat s}\right)\right]\;; \label{eq:real_ham1}\\
&\frac{\dmat (\im q)}{\dmat s} = \phantom{-}\frac{\partial}{\partial (-\im p)} \left[\re \left(H\frac{\dmat t}{\dmat s}\right)\right]\;; \label{eq:real_ham2}\\
&\frac{\dmat (\re p)}{\dmat s} = -\frac{\partial}{\partial (\re q)} \left[\re \left(H\frac{\dmat t}{\dmat s}\right)\right]\;; \label{eq:real_ham3}\\
&\frac{\dmat (-\im p)}{\dmat s} = -\frac{\partial}{\partial (\im q)} \left[\re \left(H\frac{\dmat t}{\dmat s}\right)\right]\;, \label{eq:real_ham4}
\end{align}
\end{subequations}
that describes the evolution of a system with the four real canonical 
variables $(\re p, -\im p)$ and $(\re q, \im q)$ under the Hamiltonian 
$\re [H(\dmat t/\dmat s)]$.

\begin{figure}[t]
\center \includegraphics[width=0.5\textwidth]{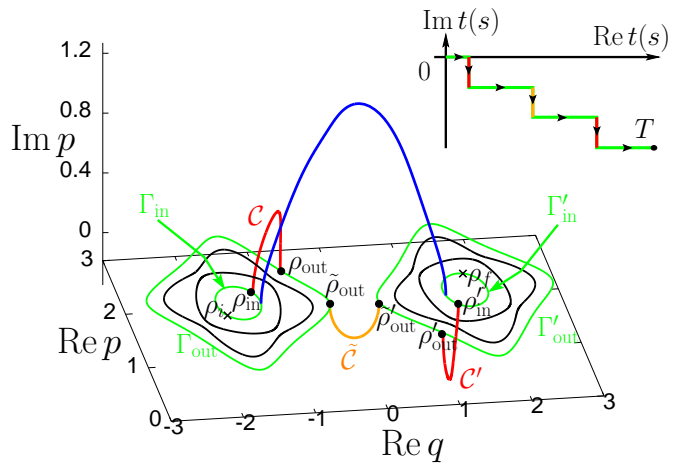}
\caption{\label{fig:ps_def}
  (Color online) Phase space of the Hamiltonian \eqref{eq:H4model} (thin black lines)
  for $a_1=1$, $a_2=-0.55$, $|b|=0.05$, and $\phi=\pi/4$,
  plotted together with two different complex trajectories 
  at energy $E\simeq 0.035$ that connect an arbitrarily chosen initial point 
  $\rho_i\equiv(p_i,q_i)$ on the real torus $\Gamma_{\goin}$ with an 
  arbitrarily chosen final point $\rho_f\equiv(p_f,q_f)$ on the
  symmetric counterpart $\Gamma'_{\goin}$ [both tori are plotted in (light) green].
  The (red and orange) arcs are half of the complex orbits (\textit{ii}) 
  $\loopC$, $\loopC'$, and (\textit{iv}) $\loopCtilde$ described in the text.
  Together with the connecting pieces of $\Gamma_{\goou}$ and
  $\Gamma'_{\goou}$, they constitute a complex trajectory that results from
  the time path depicted in the upper right inset.
  The (dark) blue line is a complex orbit lying on a part of the complex manifold
  that directly connects the symmetric tori $\Gamma_{\goin}$ and
  $\Gamma'_{\goin}$ (see Fig.~\ref{fig:ps_modif}).
  These two trajectories are plotted in a reduced three-dimensional
  complex phase space spanned by $(\re p, \re q, \im p)$. 
}
\end{figure}

In our case of resonance-assisted tunneling, we start from an 
initial point $(p_i,q_i)\equiv\big(p(s_i),q(s_i)\big)$ at time 
$t_i\equiv t(s_i)$ on a real inner torus $\Gamma_{\goin}$ inside the eye 
of one of the two main islands and choose a time path $t(s)$ with the 
shape of a descending staircase as sketched in Fig.~\ref{fig:ps_def}. 
This time path is not restricted to lie all along the imaginary axis 
as imposed by the theory of instantons.
Instead, it successively evolves along the real and imaginary directions 
(characterized by a real and imaginary $\dmat t/\dmat s$, respectively), 
such that the complex trajectory ends, after a time $T$, at the final point
$(p_f,q_f)$ on the real torus $\Gamma'_{\goin}$ that corresponds
to the counterpart of $\Gamma_{\goin}$ in the symmetric island.
The freedom in the choice of $t(s)$ can be justified from the fact that 
the semiclassical contributions to tunneling, as obtained through
the stationary phase approximation of the time propagator $G(q_f,q_i,T)$,
arise from action integrals along complex classical trajectories that 
join $q_i$ and $q_f$ in a time $T$ and fulfill Eqs.~\eqref{eq:real_ham}. 
McLaughlin showed~\cite{McLaughlin_1972} that these integrals are 
independent of the time path $t(s)$ as long as no bifurcations of 
trajectories are encountered while deforming the path and as long as 
$\im t(s)$ does not increase with $s$ in order to guarantee the 
boundedness of any intermediate evolution operator.

Tuning properly the length of the stairs as we depict in 
Fig.~\ref{fig:ps_def}, complex trajectories then can be described as a 
continuous concatenation of pieces of the following distinct orbits:

(\textit{i}) the two symmetric real periodic orbits
lying on the inner tori $\Gamma_{\goin}$ and $\Gamma'_{\goin}$
with the real energy $E$, the real period 
$T_{\goin}(E) = T'_{\goin}(E)$, and the 
real action $S_{\goin}(E) = S'_{\goin}(E)$;

(\textit{ii}) the two symmetric complex periodic orbits 
$\loopC$ and $\loopC'$ with the imaginary period 
$\imat T_{\goc}(E)  = \imat T'_{\goc}(E)$ 
and the imaginary action $\imat\sigma_{\goc}(E) 
=\imat\sigma'_{\goc}(E)$ with $\sigma_{\goc}(E)>0$,
which connect the real inner tori
$\Gamma_{\goin}$ and $\Gamma'_{\goin}$ with the outer ones
$\Gamma_{\goou}$ and $\Gamma'_{\goou}$, respectively;

(\textit{iii}) the two symmetric real periodic orbits
lying on the outer tori $\Gamma_{\goou}$ and $\Gamma'_{\goou}$
with the real period $T_{\goou}(E) =T'_{\goou}(E)$ and the 
real action $S_{\goou}(E) =S'_{\goou}(E) > 0$;

(\textit{iv}) a complex periodic orbit $\loopCtilde$ defined
on the complex manifold that connects the outer tori
$\Gamma_{\goou}$ and $\Gamma'_{\goou}$, with the imaginary action
$\imat\tilde{\sigma}_{\goc}(E)$ with $\tilde{\sigma}_{\goc}(E) > 0$
and the imaginary period $\imat \tilde{T}_{\goc}(E)$.

\begin{figure}[t]
\center \includegraphics[width=0.45\textwidth]{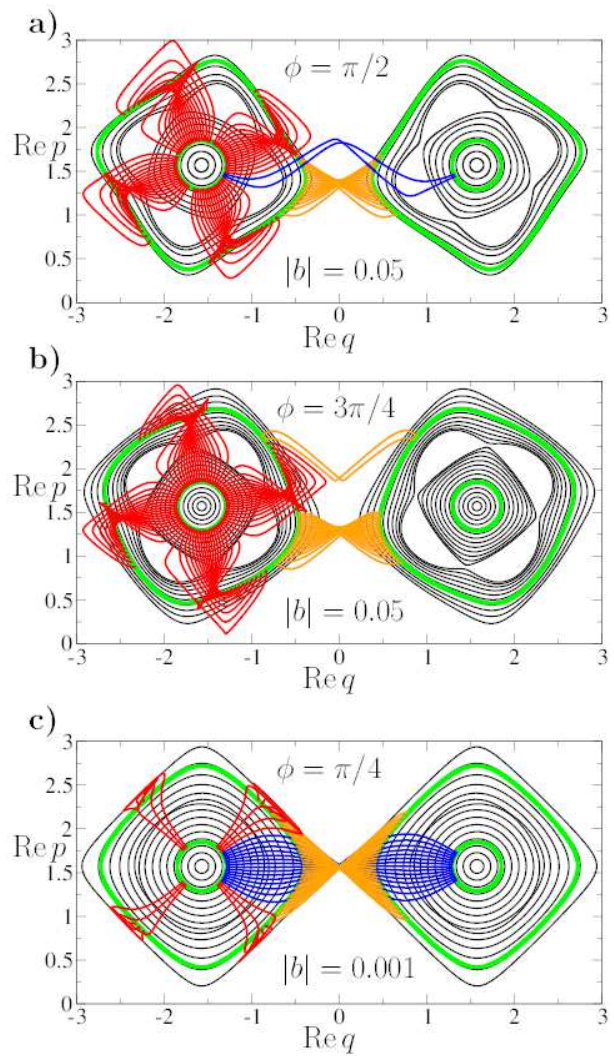}
\caption{\label{fig:ps_modif} 
(Color online) Visualization of the complex manifold associated with a pair of
inner real tori $\Gamma_{\goin}$, $\Gamma'_{\goin}$ and the corresponding 
outer real tori $\Gamma_{\goou}$, $\Gamma'_{\goou}$ at energy 
$E\simeq 0.035$ (all the four are plotted with thick green lines) 
for the Hamiltonian \eqref{eq:H4model} with $a_1=1$, $a_2=-0.55$, and 
(a) $|b|=0.05$, $\phi=\pi/2$; (b) $|b|=0.05$, $\phi=3\pi/4$; 
(c) $|b|=0.001$, $\phi=\pi/4$.
An exemplary set of complex orbits starting from different initial points on 
$\Gamma_{\goin}$ and $\Gamma_{\goou}$ is projected onto the real phase space 
(using the same color code as in Fig.~\ref{fig:ps_def}), in order to 
illustrate the topology of this complex manifold. 
While there are two distinct families of (orange) complex orbits 
that connect the outer real tori of the two islands in (b), 
only the lower family (at $p<\pi/2$) contributes to the tunneling process 
as its imaginary action is significantly smaller than the one of the upper 
family (at $p>\pi/2$).}
\end{figure}

The closed orbits (\textit{i}) -- (\textit{iv}) are geometrical
objects with the property that the values of the associated actions
do not depend on the choice of the canonical coordinates. 
Those actions are given in the $(p,q)$-representation by
\begin{subequations}
\begin{align}
S_{\goin}(E) &= \oint_{\Gamma_{\goin}}{\re p \; \dmat (\re q)}\;; 
\label{eq:action_re_in} \\
S_{\goou}(E) &= \oint_{\Gamma_{\goou}}{\re p \; \dmat (\re q)}\;; 
\label{eq:action_re_ou} \\
\sigma_{\goc}(E) &= \oint_{\loopC} 
\left[\re p \; \dmat(\im q) + \im p \; \dmat(\re q) \right]\;; 
\label{eq:action_im_c} \\
\tilde{\sigma}_{\goc}(E) &= \oint_{\tilde{\loopC}}
\left[\re p \; \dmat(\im q) + \im p \; \dmat(\re q) \right]\;. 
\label{eq:action_im_c_tilde}
\end{align}
\end{subequations}
For the two last actions, the contours can be continuously deformed 
according to Cauchy's theorem as long as no singularities of 
the transformation $q\mapsto p(q,E)$ are crossed 
(see Fig.~\ref{fig:ps_modif}).

Starting first with a portion of real time $(\dmat t/\dmat s>0)$, the
trajectory evolves from the initial point $\rho_i\equiv(p_i,q_i)$ 
along the 
torus $\Gamma_{\goin}$ until it reaches a certain point $\rho_{\goin}$. 
Then the time varies along the imaginary direction $(\imat\dmat t/\dmat s>0)$,
driving the trajectory into the complex domain until it reaches again 
the real phase space, namely at the point $\rho_{\goou}$ on the outer
torus $\Gamma_{\goou}$. The length of this time step is equal to
$|T_{\goc}(E)/2|$ such that only a half of the closed orbit $\loopC$ is
followed in order to reach the outer torus.
Another portion of real time is again spent to evolve on the real torus 
$\Gamma_{\goou}$ and reach the real point $\tilde{\rho}_{\goou}$. 
Then again, thanks to an imaginary time step with the length
$|\tilde{T}_{\goc}(E)/2|$, the trajectory evolves on the complex
manifold and joins, after a half of a loop $\loopCtilde$, the torus
$\Gamma'_{\goou}$ at a point $\tilde{\rho}'_{\goou}$ on the
other side of the main separatix delimiting the two main islands. 
By symmetry, this procedure is repeated to connect consecutively the real
phase-space points $\rho'_{\goou}$ and $\rho'_{\goin}$ (for simplicity 
they can be taken as the  symmetric partners of $\rho_{\goou}$ and 
$\rho_{\goin}$, respectively, though this is not necessary) and 
finally $\rho_f\equiv(p_f,q_f)$. 
Invoking again Cauchy's theorem, the exact location 
of $\rho_{\goin}$, $\rho_{\goou}$, $\rho'_{\goin}$, and $\rho'_{\goou}$ 
on the real plane is not important as long as no singularity is 
encountered when moving them along the corresponding real tori 
(see Fig.~\ref{fig:ps_modif})
\footnote{In principle, one would even be allowed to follow a time path 
where $\re t(s)$ decreases with $s$ somewhere along the path.
However, as the final time $T$ is fixed in the semiclassical
framework developed in Ref.~\cite{LeDeunff.Mouchet_2010}, one would, in 
this latter case, have to compensate this backward motion by the exactly 
same forward contribution on the symmetric torus in order to reach $T$.
The real and imaginary action integrals along this time path would
therefore be exactly the same as for the time path shown in 
Fig.~\ref{fig:ps_def}.}.

We now use the different parts of the generic complex trajectory we
have just described in order to split the tunneling process up into
two main steps, namely 
(I) to cross the separatrices that delimit the resonant chains and
(II) to pass over the separatrix structure that divides the two main islands.

(I) From its very construction, the global Hamiltonian
\eqref{eq:model_period_H4} can be approximated by the normal form
\eqref{eq:model_h4} in the neighborhood of~$(q,p)=(\pm\pi/2,\pi/2)$
and rewritten in action-angle coordinates using the canonical transformation 
$(p=\pi/2 + \sqrt{2I}\cos \theta,q=\pm\pi/2 + \sqrt{2I}\sin \theta)$ with 
$I>0$ and $\theta \in [0,2\pi]$.
This finally yields a modified mathematical pendulum
\begin{equation}\label{eq:pend_approx}
H_{\mathrm{loc}}(I,\theta) = K_0 + \frac{(I-I_{\ell:1})^2}{2m_{\ell:1}} + 
2V(I)\cos{(\ell\theta+\phi_{\ell:1})}\;,
\end{equation}
which is parametrized by coefficients that are given by the 
parameters of the exact global Hamiltonian:
\begin{subequations}
\begin{align}
&K_0 = -a_1^2/(16a_2)\;, \quad I_{\ell:1} = -a_1/(8a_2)\;, 
\label{eq:pend_approx_coeff1}\\
&\phi_{\ell:1} = \phi\;, \quad m_{\ell:1} = 1/(8a_2)\;, 
\quad V(I) = 2|b|I^2\;. \label{eq:pend_approx_coeff2}
\end{align}
\end{subequations}
This pendulum structure provides a dynamical tunnel coupling
between the inner torus $\Gamma_{\goin}$ and the outer torus $\Gamma_{\goou}$.

To describe this tunneling process by means of the JWKB theory, 
we represent the global quasimode $\ket{\Psi}$ localized
on the quantized torus $\Gamma_{\goin}$,
which is characterized by the energy $E_n$ and the oscillation period
$T_{\goin}(E_n)$ in angle representation as~\cite[Appendix C]{Brodier.Schlagheck.ea_2002}
\begin{equation}
\Psi(\theta) \simeq \frac{1}{\sqrt{T_{\goin}(E_n)|\dot{\theta}|}}\left[ \EXP{\imat \mathcal{S}_{\goin}(\theta,E_n)/\hbar} + \coeffAT \EXP{\imat [\mathcal{S}_{\goou}(\theta,E_n)/\hbar + \mu]} \right]
\end{equation}
where $\dot{\theta}$ denotes the time derivative of the angle coordinate.
The additional phase $\mu$ comes from the consistency with
Langer connection formulas~\cite{Langer_1937} and is related to the
Maslov index counting the number of caustics encountered along the
classical trajectory with the corresponding real action
$\mathcal{S}_{\goin}(\theta,E_n) = \int_{0}^{\theta}{I(\theta',E_n)\dmat\theta'}$ 
where $I(\theta',E_n)$ indicates the action coordinate
along the torus $\Gamma_{\goin}$. 
By construction, the action over a period, which is given by 
Eq.~\eqref{eq:action_re_in}, is quantized according to 
$S_{\goin}(E_n) \equiv \mathcal{S}_{\goin}(2\pi,E_n) = 2\pi\hbar(n+1/2)$. 
On the other hand, the torus $\Gamma_{\goou}$ on the outer side of the 
resonance chain, with the action $S_{\goou}(E_n)$ given by 
Eq.~\eqref{eq:action_re_ou}, is not \textit{a priori} quantized and thus 
$S_{\goou}(E_n)/(2\pi\hbar)-1/2$ is not an integer in general. 
The coupling amplitude $\coeffAT$, which characterizes the 
tunneling-induced admixture of the component associated with 
$\Gamma_{\goou}$ to the quasimode on the inner torus 
$\Gamma_{\goin}$, then can be evaluated as~\cite{Brodier.Schlagheck.ea_2002}
\begin{equation}\label{eq:amp_trans}
\coeffAT = \frac{\EXP{-\sigma_{\goc}(E_n)/(2\hbar)}}{2\sin{\left[(S_{\goin}(E_n)-S_{\goou}(E_n))/(2\ell\hbar)\right]}},
\end{equation}
where the half of the imaginary action $i \sigma_{\goc}(E_n)$
of the closed loop $\loopC$ defined by~\eqref{eq:action_im_c} is involved.

(II) We now make use of the part of the trajectory that connects 
the outer torus $\Gamma_{\goou}$, which has the same energy $E=E_n$ 
as $\Gamma_{\goin}$, to its symmetric counterpart $\Gamma_{\goou}'$ 
in the other cell. 
Replacing within Eq.~\eqref{eq:splitting_unpert} $\Sigma$ by the action 
of the closed orbit $\tilde{\loopC}$ and $\omega_{n}$ by the frequency 
$\omega_{\goou}$ of the outer tori, and taking into account the 
periodicity of our system (which gives rise to an additional factor two 
in the splitting formula as compared to simple double-well tunneling),
we evaluate the level splitting associated with the tori $\Gamma_{\goou}$ 
and $\Gamma'_{\goou}$ at the energy $E$ due to direct tunneling 
across the main separatrix as
\begin{equation}\label{eq:splitting_last_step}
\delta E(E) = \frac{2\hbar\omega_{\goou}}{\pi}
\EXP{-\tilde{\sigma}_{\goc}(E)/(2\hbar)}.
\end{equation}

Collecting the results~\eqref{eq:amp_trans} and \eqref{eq:splitting_last_step},
we obtain as a key statement of our paper the semiclassical prediction
\begin{equation}\label{eq:splitting_cmplx_path}
\Delta E_n = |\coeffAT|^2\delta E(E_n)
\end{equation}
for the level splitting associated with the states $\ket{\phi_n^\pm}$
\cite{Brodier.Schlagheck.ea_2002}. 
By symmetry, one needs to take into account twice the first step, 
leading to the square of the transmitted amplitude $\coeffAT$. 

\subsection{Comparison and discussion} \label{subsec:comp}

\begin{figure}[t]
\center \includegraphics[width=0.5\textwidth]{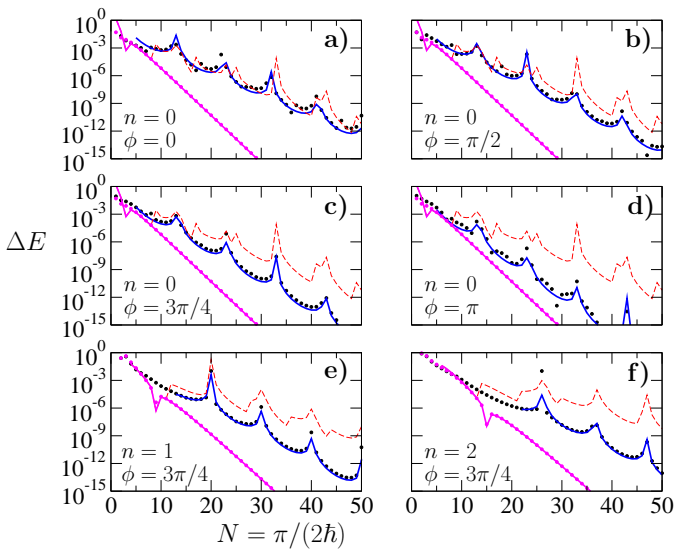}
\caption{\label{fig:split_phivar} 
(Color online) Quantum and semiclassical level splittings
$\Delta E_n$ plotted in a semilogarithmic scale versus the integer 
$N \equiv \pi/(2\hbar)$ for the Hamiltonian \eqref{eq:H4model} with 
$a_1=1$, $a_2=-0.55$, $|b|=0.05$ for the following phases and levels 
(a) $\phi = 0$, $n = 0$; (b) $\phi = \pi/2$, $n = 0$;
(c) $\phi = 3\pi/4$, $n = 0$; (d) $\phi = \pi$, $n = 0$;
(e) $\phi = 3\pi/4$, $n = 1$; (f) $\phi = 3\pi/4$, $n = 2$.
The (black) dots represent the exact numerical results while the 
(blue) solid lines show the predictions obtained by the semiclassical 
formula~\eqref{eq:splitting_cmplx_path} \cite{rem_sep}.
The (red) dashed line is the perturbative RAT prediction obtained with 
the expression~\eqref{eq:splitting_RAT} which does not depend on $\phi$.
The diagonal straight lines (plotted in magenta) correspond to the
unperturbed semiclassical prediction~\eqref{eq:splitting_unpert} while 
the (magenta) dots on top show the exact splittings for the case $b=0$. 
The dips in the unperturbed splittings around $N\simeq 3, 9, 15$, 
respectively, for $n=0, 1,2$ arise when the unperturbed quantized torus 
is located right on the crown of the volcanos. 
In that case, the classical frequency of the torus vanishes and, 
as a consequence, Eq.~\eqref{eq:splitting_unpert} predicts a vanishing 
level splitting.}
\end{figure}

\begin{figure}[t]
\center \includegraphics[width=0.5\textwidth]{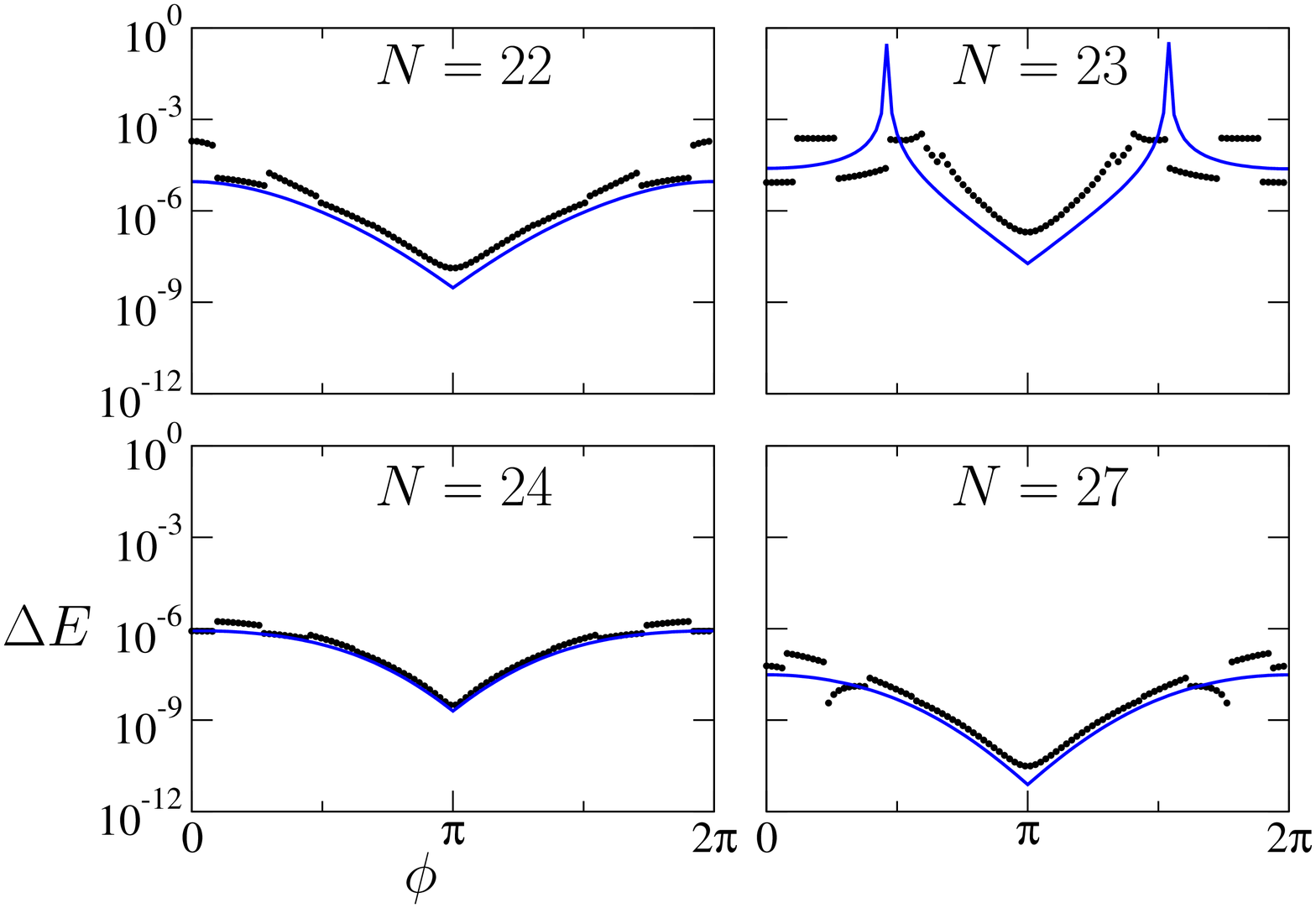}
\caption{\label{fig:split_phivar2} 
(Color online) Quantum and semiclassical level splittings 
$\Delta E_0$ associated with the eigenstates that are most strongly 
localized on the centres of the four symmetric islands, for the 
Hamiltonian \eqref{eq:H4model} with $a_1=1$, $a_2=-0.55$, $|b|=0.05$. 
The splittings are plotted versus the phase $\phi$ for different
values of $N\equiv\pi/(2\hbar) = 22$ (upper left panel),
$23$ (upper right panel), $24$ (lower left panel),and $27$ (lower right panel).
The (black) dots represent the exact numerical results
while the (blue) solid lines are the semiclassical predictions 
obtained from Eq.~\eqref{eq:splitting_cmplx_path}.}
\end{figure}

A comparison of the formula \eqref{eq:splitting_cmplx_path} with 
the exact splittings, which are obtained through numerical
diagonalization, yields a very good agreement, as shown in
Figs.~\ref{fig:split_phivar} and \ref{fig:split_phivar2}.
Peaks appear in the splitting whenever the denominator of
$\coeffAT$ vanishes, that is to say, when $\Gamma_{\goou}$
is a EBK quantized torus with a quantum number $\tilde{n}$ that satisfies
$\tilde{n} = n + \nu \ell$ with integer $\nu$.
In that case, the area $S_{\goou}(E_n)-S_{\goin}(E_n)$
enclosed by the two tori $\Gamma_{\goin}$ and $\Gamma_{\goou}$
corresponds to exactly $\nu\ell$ Planck cells of size $2\pi\hbar$.

For finite values of the perturbation strength $|b|$, the rotation angle 
$\phi$ of the classical resonant chains clearly influences the splittings,
which exhibit a symmetry axis at $\phi=\pi$ as shown in 
Fig.~\ref{fig:split_phivar2}.
Indeed, this behavior can be explained in terms of complex paths.
Keeping in mind the local pendulum approximation~\eqref{eq:pend_approx}, 
the real tori and the complex orbits $\loopC,\loopC'$ that cross
the resonance chains are not appreciably affected by a variation of 
$\phi$ which essentially corresponds to a rotation
of the resonance structures in the phase space. Correspondingly,
the peaks observed in the splittings remain globally at the same 
position when $\phi$ is varied (although they may be shifted a bit, 
see the upper right panel of Fig.~\ref{fig:split_phivar2}).
On the other hand, the imaginary action of the orbit $\tilde{\loopC}$ 
that crosses the main separatrix between the islands
is significantly modified under variation of $\phi$.
As one can indeed see in Fig.~\ref{fig:ps_modif}, the complex ``bridge''
(plotted in orange) that connects the two outer symmetric tori of the 
main islands is shifted farther away from the horizontal symmetry axis as 
$\phi$ is increased, which is naturally accompanied by an increase of the 
corresponding imaginary action. 
This is responsible for the drastic decrease of the splitting 
(by three orders of magnitude for $N\sim25$ as is seen in 
Fig.~\ref{fig:split_phivar2}) as $\phi$ is varied from 0 to $\pi$. 

It is instructive to compare the exact splittings and their semiclassical
prediction also with the perturbative theory of resonance-assisted 
tunneling (RAT), which was first introduced for 1D time-periodic 
Hamiltonians in the quasi-integrable regime
\cite{Brodier.Schlagheck.ea_2001,Brodier.Schlagheck.ea_2002}
and later extended to mixed regular-chaotic systems 
\cite{Eltschka.Schlagheck_2005,Lock.Backer.ea_2010,Lock.Backer.ea_2012}.
Following the derivation described in the Appendix~\ref{app:RAT}, 
the level splitting associated with the eigenstates $\ket{\phi_n^{\pm}}$ 
of the Hamiltonian is given by 
\begin{equation}\label{eq:splitting_RAT}
\Delta E_{n} = \Delta E^{(0)}_{n} + 
\sum_{k>0}^{k_c}{|\coeffB_{n,k\ell}|^2\Delta E^{(0)}_{n+k\ell}}\;,
\end{equation}
with $\ell\equiv4$  and
\begin{eqnarray}\label{eq:psi_approx_coeff}
\coeffB_{n,k\ell} &=& \prod_{p=1}^{k}
{\frac{A_{n+p\ell,n+(p-1)\ell}}{E_{n}^{(0)}-E_{n+p\ell}^{(0)}}}\;; \\
A_{n+p\ell,n+(p-1)\ell} &=& 2|b|\EXP{\imat\phi}\hbar^{p\ell/2}
\sqrt{\frac{(n+p\ell)!}{[n+(p-1)\ell]!}}\;,
\end{eqnarray}
where the $E^{(0)}_ {n+p\ell}$ denote the unperturbed energies (i.e.\ for $b=0$) 
and
\begin{equation}
\Delta E^{(0)}_{n+k\ell} \simeq \frac{2 \hbar\omega^{(0)}_{n+k\ell}}{\pi}
\EXP{-\sigma^{(0)}_{n+k\ell}/2\hbar}
\label{eq:DE02}
\end{equation}
the unperturbed splittings. 
The latter are determined through a numerical evaluation of the 
oscillation frequencies $\omega^{(0)}_{n+k\ell}$ and the imaginary actions 
$\sigma^{(0)}_{n+k\ell}$ associated with the unperturbed invariant tori 
at energy~$E_0$.
Using the quadratic approximation \eqref{eq:harmonic_approx}, one notices 
that the denominators of the coefficients \eqref{eq:psi_approx_coeff} are 
proportional to the quantized action variables of the unperturbed system: 
$(E_{n}^{(0)}-E_{n+p\ell}^{(0)}) \propto 
(I_{n}-I_{n+p\ell})(I_{n}+I_{n+p\ell}-2I_{\ell :s})$. 
The coupling between the unperturbed states $\ket{n}$ and $\ket{n+p\ell}$ 
is therefore maximized when the quantized tori $I_{n}$ and $I_{n+p\ell}$ 
are symmetrically located with respect to the resonant island chain 
which is approximately localized at $I_{\ell :s}$. 
This then leads to significant local enhancements of the splittings. 

However, the above formulation of the perturbative RAT theory does
not account for a modification of the imaginary actions 
$\sigma^{(0)}_{n+k\ell}$ due to the presence of the resonance chain.
While this modification can be safely neglected in generic near-integrable
systems which generally exhibit perturbatively small resonance chains
\cite{Brodier.Schlagheck.ea_2001,Brodier.Schlagheck.ea_2002}, it does 
matter in our special case of integrable resonance-assisted tunneling 
with macroscopically large resonance islands as we pointed out above.
We are therefore already beyond the perturbative regime.
This is clearly seen in Fig.~\ref{fig:split_phivar}:
Even though the perturbative RAT predictions are, by coincidence, 
still in approximate agreement with the exact splittings for $\phi = 0$ 
\footnote{We attribute this coincidence to the fact that the
complex orbit $\loopCtilde$ connecting the two symmetric outer tori
can, for $\phi=0$, still be defined along the horizontal symmetry axis 
$\re p=\pi/2$ as in the unperturbed case $b=0$.},
they drastically overestimate the latter for $\phi = \pi$.

\begin{figure}[t]
\center \includegraphics[width=0.5\textwidth]{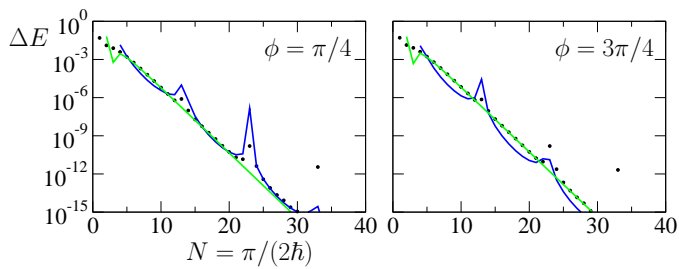}
\caption{\label{fig:split_phivar_small_b} 
(Color online) Quantum and semiclassical level splittings $\Delta E_0$
associated with the eigenstates that are most strongly localized on
the centers of the four symmetric islands, for the Hamiltonian 
\eqref{eq:model_period_H4} with the parameters $a_1=1$, $a_2=-0.55$ and 
$|b|=0.001$, plotted versus $N\equiv\pi/(2\hbar)$ for $\phi=\pi/4$ 
(left panel) and $\phi=3\pi/4$ (right panel). 
The (black) dots represent the exact results, the dark (blue) solid lines 
correspond to the resonance-assisted semiclassical splittings obtained 
with the semiclassical formula~\eqref{eq:splitting_cmplx_path}, and the 
light (green) straight lines show the direct splittings given by 
Eq.~\eqref{eq:splitting_direct}.
The effect of the resonance is significantly reduced as compared to 
Fig.~\ref{fig:split_phivar}, and the splittings no longer display
systematic variations as a function of the rotation angle $\phi$.}
\end{figure}

Let us finally discuss the interplay of resonance-assisted tunneling
with direct tunneling in the deep perturbative regime.
With decreasing $|b|$, the splittings are less and less 
sensitive to a variation of the phase $\phi$ and the resonance
peaks become less and less pronounced.
Direct tunneling becomes the dominant mechanism in the limit 
$|b|\to0$, and the splittings display a purely exponential decrease
with $1/\hbar$, which can be evaluated as
\begin{equation}\label{eq:splitting_direct}
\delta E_n^{(d)} = \frac{2\hbar\omega_{\goin}}{\pi}\EXP{-\Sigma(E_n)/(2\hbar)},
\end{equation}
where $\imat\Sigma(E_n)$ [with $\Sigma(E_n)>0$] is the imaginary action 
of the complex manifold (plotted in blue in Fig.~\ref{fig:ps_modif}) 
that directly connects the real tori $\Gamma_{\goin}$ and $\Gamma'_{\goin}$, 
and $\omega_{\goin}$ is the frequency of those two tori.
In the limit $|b| \to 0$, the action $\imat\Sigma(E_n)$ 
coincides with the unpertubed one $\imat\Sigma^{(0)}(E_n)$ and, hence, 
$\delta E_n^{(d)}$ approaches the unperturbed splittings $\Delta E_n^{(0)}$.

In the context of the RAT theory, 
L\"ock \textit{et al}~\cite{Lock.Backer.ea_2010} developed a quantitative
criterion providing, for a given strength of the perturbation, 
the characteristic value $\hbar_{\text{res}}$ of Planck's constant
that separates the ``direct'' regime ($\hbar > \hbar_{\text{res}}$)
in which direct tunneling dominates from the ``resonance-assisted'' 
regime ($\hbar < \hbar_{\text{res}}$) in which the relevant mechanism is 
resonance-assisted tunneling. 
This criterion is given by the equality
\begin{equation}\label{eq:criterion}
\frac{(\Delta\coeffA_{\ell :s})^2}{\ell\hbar_{\text{res}}\coeffA_{\ell :s}}\sqrt{\frac{\Delta E^{(0)}_{n+\ell}(\hbar_{\text{res}})}{\Delta E^{(0)}_{n}(\hbar_{\text{res}})}}\frac{1}{(\hbar_{\text{res}}/\hbar_{\text{peak}})-1} = \frac{256}{\pi}\;,
\end{equation}
where $\Delta\coeffA_{\ell :s} = 16\sqrt{2m_{\ell :s}V_{\ell :s}}$ 
is the area covered by the island chain~\cite{Eltschka.Schlagheck_2005} 
as derived from the pendulum approximation~\eqref{eq:pend_approx},
$\coeffA_{\ell :s} = 2\pi I_{\ell :s}$ is the area enclosed by the approximate 
resonant torus of the unperturbed system, and
$\hbar_{\text{peak}}$ is the value of Planck's constant at 
which the first peak in the splittings $\Delta E_n$ appears. 
Using $\hbar_{\text{peak}} \equiv \pi / ( 2 N_{\text{peak}} )$ with 
$N_{\text{peak}} \simeq 13.5(, 20, 26.25)$ for the level 
$n=0(, 1, 2)$, as extracted from Figs.~\ref{fig:split_phivar}
and \ref{fig:split_phivar_small_b}, we obtain
$N_{\text{res}} \equiv \pi/(2\hbar_{\text{res}}) \simeq 9(, 12, 14)$ 
for $|b|=0.05$ and $N_{\text{res}} \equiv \pi/(2\hbar_{\text{res}}) \simeq 13$ 
for $|b|=0.001$, which is in rather good agreement with 
Figs.~\ref{fig:split_phivar} and \ref{fig:split_phivar_small_b}, respectively.

\section{Conclusions}

In summary, we discussed a semiclassical theory of 
resonance-assisted tunneling in integrable systems, which is based
on the analytic continuation of the invariant classical tori of the system
to the complex domain.
To this end, we showed how to construct a class of 1D integrable 
Hamiltonians, based on the normal form theory, that exhibit islands of 
bounded motion surrounded by chains that mimic the resonance structures 
arising in Poincar\'e sections of nonintegrable systems. 
We then studied tunneling between two symmetric islands 
in such integrable systems.
Our semiclassical theory, which is essentially expressed by 
Eqs.~\eqref{eq:amp_trans} and \eqref{eq:splitting_cmplx_path}, 
is found to reproduce the numerically computed tunneling splittings with
rather good accuracy.
In contrast to the standard implementation of the RAT theory which is based
on quantum perturbation theory, Eqs.~\eqref{eq:amp_trans} and 
\eqref{eq:splitting_cmplx_path} provide reliable predictions of level 
splittings also in the nonperturbative regime characterized by rather 
well-developed resonance island chains.
In that case, a rotation of the resonance chain with respect to the 
main separatrix of the system may have a significant impact on the 
tunneling rates of the system, due to the associated displacement 
of the invariant manifolds that cross the separatrix and govern
direct tunneling outside the resonance chain.

Even though a full derivation of our semiclassical theory is not 
presented here, the similarity of Eqs.~\eqref{eq:amp_trans} and 
\eqref{eq:splitting_cmplx_path} with the analytical expression for 
the level splittings in a triple-well potential derived in 
Ref.~\cite{LeDeunff.Mouchet_2010} [see Eq.~(66) there] suggests 
that our main result~\eqref{eq:splitting_cmplx_path} could eventually 
be derived by adapting the semiclassical framework developed in 
Ref.~\cite{LeDeunff.Mouchet_2010} [see Eq.~(40) there]
to our case of resonance-assisted tunneling through island chains
(although this latter case topologically differs markedly from the
triple-well system).
Indeed, we expect that the resonance peaks observed in the 
splittings arise due to constructive Fabry-P\'erot type
interferences between topologically distinct complex trajectories 
that connect the two islands via the periodic orbits introduced 
in Sec.~\ref{subsec:semiclassical_formula}, which are frequented 
with distinct repetitions.
The main difficulty in establishing such a semiclassical framework 
in the spirit of Ref.~\cite{LeDeunff.Mouchet_2010} comes from the 
selection of the class of complex orbits that brings the main contribution. 
For a Hamiltonian of the form~\eqref{eq:usual_ham}, the restriction to
complex periodic orbits along which one canonical variable remains purely 
real provides a useful help. 
However, an analogous restrictive criterion is not known to us for the 
resonance-assisted tunneling problem under consideration.

Through recursive application of the basic principle of resonance-assisted
tunneling, one may expect to derive a generalized semiclassical expression
for level splittings between islands that contain $R>1$ different 
$(\ell_r: s_r)$ island chains for~$r=1,\dots,R$. 
Such a semiclassical expression is expected to be of the form
\begin{equation}\label{eq:split_generalized}
\Delta E_n = \left[\prod_{r=1}^{R}{\big|\coeffAT^{(\ell_r: s_r)}(\hbar)\big|^2}\right]\delta E(E_n)\;,
\end{equation}
where $\delta E(E_n)$ is the direct splitting~\eqref{eq:splitting_last_step} 
associated with the outermost torus involved in this multiresonance 
transition process.
In the perturbative regime, the individual coupling amplitudes
$\coeffAT^{(\ell_r: s_r)}(\hbar)$ can be approximately 
evaluated using a local pendulum approximation for each 
$(\ell_r: s_r)$ resonance
chain (see also Eq.~(76) in Ref.~\cite{Brodier.Schlagheck.ea_2002}).
However, there is no guarantee that this approach remains valid 
in the presence of nonperturbatively large resonance chains. 
A careful investigation of the complex manifolds will be required 
in that case in order to determine which type of complex paths are 
relevant depending on the relative size of the chains with 
respect to Planck's constant.

Finally, our theory may provide a useful starting point for developing a
quantitative semiclassical description of tunneling also in nonintegrable
systems that exhibit a mixed regular-chaotic phase space structure.
In analogy with the perturbative RAT study of Ref.~\cite{Lock.Backer.ea_2010},
resonance-assisted transitions will, in that case, have to be combined with
direct regular-to-chaotic tunneling \cite{Backer.Ketzmerick.ea_2008}
for which a fully semiclassical theory in terms of complex paths was
recently presented in Ref.~\cite{Mertig.Lock.ea_2013}.
It seems straightforward to incorporate the effect of nonlinear resonances
into this latter semiclassical framework of Ref.~\cite{Mertig.Lock.ea_2013},
in order to extend its applicability to the deep semiclassical regime in which
resonances generically play a role.

\section*{Acknowledgments}

This work was financially supported by the Bayerisch-Franz\"osisches 
Hochschulzentrum (BFHZ-CCFUB), by the Deutsche Forschungsgemeinschaft 
in the framework of the DFG Forschergruppe FOR760, and by the Advanced Study 
Group (ASG) ``Towards a Semiclassical Theory of Dynamical Tunneling'' which 
took place at the Max-Planck-Institut f\"ur Physik Komplexer Systeme (MPIPKS).
We thank the members and invited speakers of the ASG, in particular 
Roland Ketzmerick, Steffen L\"ock, Normann Mertig, Akira Shudo, and 
Denis Ullmo, for useful discussions.

\appendix
\section{Theory of resonance-assisted tunneling for 1D
integrable systems}\label{app:RAT}

In this appendix, we review the main steps of the perturbative approach 
to describe resonance-assisted tunneling (RAT) applied to the simple case of 
one-dimensional integrable Hamiltonians. In analogy with the procedure 
described in Ref.~\cite{Keshavamurthy.Schlagheck} (see also 
Refs.~\cite{Brodier.Schlagheck.ea_2001,Brodier.Schlagheck.ea_2002,Lock.Backer.ea_2010}), we start with a 1D time-independent Hamiltonian $H(p,q)$ 
(which is assumed to be an analytical function in $p$ and $q$) 
that exhibits in the phase space two main symmetric regions, 
each of them surrounded by one $\ell$:$s$ resonant island chain. 
Approximate action-angle variables $(I,\theta)$, which result from
$(p,q)$ via a canonical transformation, can be defined locally within each
of the two regions. Their time evolution is governed by a Hamiltonian of the 
form
\begin{equation}
H(I,\theta) = H_0(I) + V(I,\theta)
\end{equation}
[for the sake of simplicity, we shall keep the same notation 
$H(\cdot,\cdot)$ for both the $(p,q)$ and the $(I,\theta)$ 
representations],
where the angle-dependent perturbation $V(I,\theta)$
is responsible for the generation of the resonant chain. 
We now expand the perturbation as a Fourier series
\begin{equation}\label{eq:pert_fourier}
V(I,\theta) = \sum_{j=1}^{\infty}{2V_j(I)\cos{(j\ell\theta+\phi_j)}}\;,
\end{equation}
and perform a harmonic approximation of the angle-independent part 
of the Hamiltonian near the resonant chain,
\begin{equation}\label{eq:harmonic_approx}
H_0(I) \simeq H_0(I_{\ell :s}) + \frac{(I-I_{\ell :s})^2}{2m_{\ell :s}}
 + \Omat[(I-I_{\ell :s})^3]\;,
\end{equation}
where $I_{\ell :s}$ is the action variable at the resonance and 
$1/m_{\ell :s} \equiv \dmat^2 H_0/\dmat I^2$ at $I=I_{\ell :s}$. 
By definition, the frequency of oscillations 
$\Omega \equiv \dmat H_0/\dmat I$ vanishes at the resonance. 
Combining Eqs.~\eqref{eq:pert_fourier} and \eqref{eq:harmonic_approx} 
[and omitting the constant $H_0(I_{\ell :s})$], 
$H(I,\theta)$ is reduced to a modified generalized pendulum of the form
\begin{equation}\label{eq:gen_pendul}
H_{\mathrm{pend}}(I,\theta) \DEF \frac{(I-I_{\ell :s})^2}{2m_{\ell :s}} + \sum_{j=1}^{\infty}{2V_j(I)\cos{(j\ell\theta+\phi_j)}}\;.
\end{equation}

Treating $V(I,\theta)$ as a small perturbation, one can now apply the 
time-independent quantum perturbation theory. 
Using the eigenstates $\ket{n}$ (with $n \in\mathbb{N}_0$) 
of the operator $\hat{I} = -\imat\hbar\partial/\partial \hat{\theta}$ 
which fulfill
\begin{eqnarray}
\hat{I}\ket{n} &=& I_n\ket{n} = \hbar(n+1/2)\ket{n}\;, \\
H_0(\hat{I})\ket{n} &=& E_n^{(0)}\ket{n}\;,
\end{eqnarray}
one can notice that $\hat{V}(\hat{I},\hat{\theta})$ induces only couplings between the unperturbed states $\ket{n}$ and $\ket{n+j\ell}$ through the matrix elements 
\begin{equation}\label{eq:mat_elemnt}
A_{n+j\ell,n} \DEF \bra{n+j\ell}H_{\mathrm{pend}}(\hat{I},\hat{\theta})\ket{n}\;.
\end{equation}
In this basis, the true eigenstates of $H_{\mathrm{pend}}(\hat{I},\hat{\theta})$ can be approximated by the following expression
\begin{multline}\label{eq:state_q_pert}
\ket{\Psi_n} \simeq \ket{n} + \sum_{k}{\frac{A_{n+k\ell,n}}{E_n^{(0)}-E_{n+k\ell}^{(0)}}\ket{n+k\ell}} \\
 + \sum_{k,k'}{\frac{A_{n+k\ell,n+k'\ell}}{E_n^{(0)}-E_{n+k\ell}^{(0)}}\frac{A_{n+k'\ell,n}}{E_n^{(0)}-E_{n+k'\ell}^{(0)}}\ket{n+k\ell}} + \cdots \;.
\end{multline}
As the perturbation is analytic, one can safely assume that the 
coefficients $V_j(I)$ decrease exponentially with $j$
\cite{Brodier.Schlagheck.ea_2002}. This property is used in 
Ref.~\cite{Brodier.Schlagheck.ea_2002} to show that the
coupling between $\ket{n}$ and $\ket{n+k\ell}$ via the $k$-steps process 
involving the matrix elements $A_{n+\ell,n}$, $A_{n+\ell,n+2\ell}$, \ldots, 
$A_{n+(k-1)\ell,n+k\ell}$ is generally much stronger than the direct coupling 
via the matrix element $A_{n+k\ell,n}$. This allows us thus to retain only 
the first term $j=1$ in Eq.~\eqref{eq:gen_pendul} and to reduce the 
Hamiltonian to a modified mathematical pendulum
\begin{equation}
H(I,\theta) \simeq \frac{(I-I_{\ell :s})^2}{2m_{\ell :s}} + 2V_1(I)\cos{(\ell\theta+\phi_1)}\;.
\end{equation}
The perturbed eigenstates thus can be expressed as
\begin{equation}\label{eq:psi_approx_app}
\ket{\Psi_n} \simeq \ket{n} + \sum_{k>0}{\coeffB_{n,k\ell} \ket{n+k\ell}}\;,
\end{equation}
with
\begin{equation}\label{eq:psi_approx_coeff_app}
\coeffB_{n,k\ell} = \prod_{p=1}^{k}{\frac{A_{n+p\ell,n+(p-1)\ell}}{E_n^{(0)}-E_{n+p\ell}^{(0)}}}\;.
\end{equation}

The next step is to evaluate the coefficients $V_j(I)$. 
Using the analyticity of $H(p,q)$, one can define a local canonical 
transformation $(p,q) \mapsto (P,Q)$ such that 
$\exp{(\pm\imat j\theta)} = [(Q \mp \imat P)/\sqrt{2I}]^j$. 
In the new coordinates, the perturbation \eqref{eq:pert_fourier} reads
\begin{equation}
V(P,Q) = \sum_{j=1}^{\infty}{\frac{V_j(I)}{(2I)^{j\ell/2}}[(Q-\imat P)^{j\ell}\EXP{\imat\phi_j} + (Q+\imat P)^{j\ell}\EXP{-\imat\phi_j}]}\;,
\end{equation}
where the $V_j(I)$ must be at least of order $I^{j\ell/2}$ to be consistent with the normal form theory described in Sec.~\ref{subsec:theory}. Making the assumption that $V_j(I) = v_j I^{j\ell/2}$, we obtain
\begin{equation}\label{eq:pert_PQ}
V(P,Q) = \sum_{j=1}^{\infty}{\frac{v_j}{2^{j\ell/2}}[(Q-\imat P)^{j\ell}\EXP{\imat\phi_j} + (Q+\imat P)^{j\ell}\EXP{-\imat\phi_j}]}\;.
\end{equation}
The corresponding quantum operators 
$(\hat{P},\hat{Q})$ can be expressed in terms of the ladder operators 
$(\hat{a},\hat{a}^{\dag})$ that are associated with the eigenstates $\ket{n}$ through the relations
\begin{eqnarray}
\hat{a} &=& \frac{1}{\sqrt{2}}(\hat{Q}+ \imat \hat{P})\;; \\
\hat{a}^{\dag} &=& \frac{1}{\sqrt{2}}(\hat{Q}- \imat \hat{P})\;.
\end{eqnarray}
Using these relations in the quantization of the perturbation
\eqref{eq:pert_PQ}, the matrix elements \eqref{eq:mat_elemnt} become finally
\begin{equation}\label{eq:mat_elemnt_nf}
A_{n+j\ell,n} = v_j\hbar^{j\ell/2}\EXP{\imat\phi_j}\sqrt{\frac{(n+j\ell)!}{n!}}\;.
\end{equation}
In our case, this expression does not represent an approximation since 
the action dependence $V_j(I) = v_j I^{j\ell/2}$ of the perturbation is, 
as shown Sec.~\ref{subsec:model}, imposed by the construction procedure 
of the Hamiltonian through the framework of normal forms.

Coming back to the original Hamiltonian $H(\hat{p},\hat{q})$, 
one can define in each symmetric well two distinct quasimodes 
$\ket{\Psi^{L}_n}$ and $\ket{\Psi^{R}_n}$ that are constructed on the 
symmetric tori with the energy $E_n$. 
This energetic degeneracy is lifted due to tunneling between the wells, 
and the level splitting $\Delta E_n$ is evaluated as the coupling 
matrix element between the quasimodes: 
$\Delta E_n = \bra{\Psi^{L}_n}H(\hat{p},\hat{q})\ket{\Psi^{R}_n}$. 
Using Eqs.~\eqref{eq:psi_approx_app}, \eqref{eq:psi_approx_coeff_app} and 
\eqref{eq:mat_elemnt_nf}, the splitting for an arbitrary level can finally 
be written as
\begin{equation}
\Delta E_{n} = \Delta E^{(0)}_{n} + \sum_{k=1}^{k_c}{|\coeffB_{n,k\ell}|^2\Delta E^{(0)}_{n+k\ell}}\;.
\end{equation}
For a standard double well system, we would have
\begin{equation}
\Delta E^{(0)}_{n+k\ell} \simeq \frac{\hbar\omega^{(0)}_{n+k\ell}}{\pi}\EXP{-\sigma^{(0)}_{n+k\ell}/\hbar} \label{eq:DE0}
\end{equation}
as the splitting for the $(n+k\ell)$-th doublet of the unperturbed system 
$H_0(\hat{p},\hat{q})$, while an additional factor 2 arises on the 
right-hand side of Eq.~\eqref{eq:DE0} in our case of a periodic array 
of wells.
$\sigma^{(0)}_{n+k\ell}$ is the action of the instanton-like trajectory 
connecting the two symmetric tori with energy $E^{(0)}_{n+k\ell}$, and 
$\omega^{(0)}_{n+k\ell}$ is the corresponding oscillation frequency. 
By construction, the RAT process may only couple quasimodes 
that are localized within the same region of regular oscillations. 
The index $k_c\ell$ labels the most highly excited state that can be 
involved in a perturbative coupling scheme starting from 
the $n$-th excited state.
Defining by $\coeffA$ the area one of those regions 
(which in Fig.~\ref{fig:ps_period_H4} corresponds to the area 
enclosed by the separatrices within each cell), we have
\begin{equation}
k_c = \left\lfloor \frac{1}{\ell}\left(\frac{\coeffA}{2\pi\hbar} 
- \frac{2n+1}{2}\right) \right\rfloor\;,
\end{equation}
where $\left\lfloor \cdot \right\rfloor$ stands for the integer part 
of a real number.

%%%%%%%%%%%%%%%%%%%%%%%%%%%%%%%%%%%%%%%%%%%%%%%%%%%%%

%%%%%%%%%%%%%%%%%%%%%%%%%%%%%%%%%%%%%%%%%%%%%%%%%%%%%%%%%


\begin{thebibliography}{10}
\bibitem{rem_sep}
The latter can be properly evaluated as soon as $N>N_{\text{s}}$ 
where $N_{\text{s}}$ corresponds to the value of $\hbar$ at which the initial 
state $\ket{\phi_n}$ is semiclassically associated with the inner separatrix 
structure of the island chain.
\end{thebibliography}
\end{document}